\documentclass[fleqn,10pt]{wlscirep}

\usepackage[utf8]{inputenc}
\usepackage[T1]{fontenc}
\usepackage{graphicx}
\usepackage{epstopdf}
\usepackage{siunitx}
\usepackage{physics}
\usepackage[capitalize]{cleveref}
\usepackage[version=4]{mhchem}
\usepackage{lineno}
\usepackage{xcolor, soul}
\usepackage[nolists]{endfloat}

\newcommand{\Rb}[1]{\textsuperscript{#1}{Rb}}

\title{A Compact Cold-Atom Interferometer with a High Data-Rate Grating Magneto-Optical Trap and a Photonic-Integrated-Circuit-Compatible Laser System}

\author[1,*]{Jongmin Lee}
\author[1]{Roger Ding}
\author[1]{Justin Christensen}
\author[1]{Randy R. Rosenthal}
\author[1]{Aaron Ison}
\author[1]{Daniel P. Gillund}
\author[1]{David Bossert}
\author[1]{Kyle H. Fuerschbach}
\author[1]{William Kindel}
\author[1]{Patrick S. Finnegan}
\author[1]{Joel R. Wendt}
\author[1]{Michael Gehl}
\author[1]{Ashok Kodigala}
\author[1]{Hayden McGuinness}
\author[1]{Charles A. Walker}
\author[1]{Shanalyn A. Kemme}
\author[1]{Anthony Lentine}
\author[2]{Grant Biedermann}
\author[1]{Peter D. D. Schwindt}

\affil[1]{Sandia National Laboratories, Albuquerque, New Mexico 87185, USA}
\affil[2]{Department of Physics and Astronomy, University of Oklahoma, Norman, Oklahoma 73019, USA}
\affil[*]{jongmin.lee@sandia.gov}

\begin{abstract}
The extreme miniaturization of a cold-atom interferometer accelerometer requires the development of novel technologies and architectures for the interferometer subsystems.
Here we describe several component technologies and a laser system architecture to enable a path to such miniaturization.
We developed a custom, compact titanium vacuum package containing a microfabricated grating chip for a tetrahedral grating magneto-optical trap (GMOT) using a single cooling beam.
In addition, we designed a multi-channel photonic-integrated-circuit-compatible laser system implemented with a single seed laser and single sideband modulators in a time-multiplexed manner, reducing the number of optical channels connected to the sensor head.
In a compact sensor head containing the vacuum package, sub-Doppler cooling in the GMOT produces \SI{15}{\micro \K} temperatures, and the GMOT can operate at a \SI{20}{\Hz} data rate. We validated the atomic coherence with Ramsey interferometry using microwave spectroscopy, then demonstrated a light-pulse atom interferometer in a gravimeter configuration for a \SI{10}{\Hz} measurement data rate and $T = \SIrange[range-units=single, range-phrase=-]{0}{4.5}{\ms}$ interrogation time, resulting in $\Delta g / g = \num{2.0e-6}$.
This work represents a significant step towards deployable cold-atom inertial sensors under large amplitude motional dynamics.
\end{abstract}
\begin{document}

\flushbottom
\maketitle
\thispagestyle{empty}

\section*{Introduction}

\begin{figure}[b!]
    \centering
	\includegraphics[width=1\columnwidth]{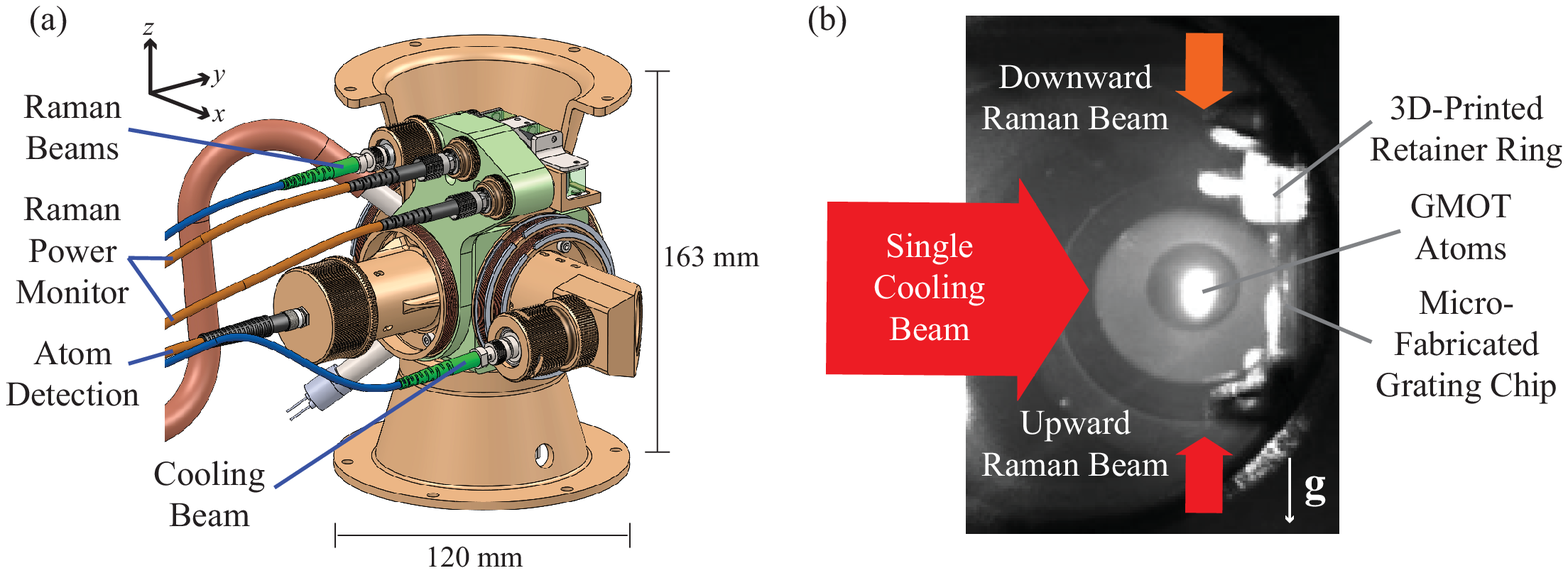}
	\caption{
			Concept of the compact light-pulse atom interferometer (LPAI) for high-dynamic conditions.
			(a) 3D rendering of the compact LPAI sensor head with fixed optical components and reliable optomechanical design.
			(b) Picture of the steady-state GMOT atoms in the sensor head.
		}
	\label{fig:sensor_head}
\end{figure}

The growing importance of advanced sensing applications \cite{Heavner14, Wynands05, Nicholson15, Ludlow15, Romalis10, Raithel14, Schwindt10, Grujic2015ASA, Maze08, Balasu08} has inspired significant investment in quantum sensing technologies. One area of particular interest is the development of cold-atom inertial sensors based on a light-pulse atom interferometer (LPAI) \cite{KasevichPRL1991, GustavsonPRL1997, Chu99, McGuirkPRA2002, CroninRMP2009, Kovachy15, ParkerScience2018}, using photon recoils from stimulated Raman transitions. The light pulse sequence of LPAIs, e.g., $\frac{\pi}{2} \rightarrow T \rightarrow \pi \rightarrow T \rightarrow \frac{\pi}{2}$, coherently splits, redirects, and recombines atomic wavepackets for matter-wave interference, where $T$ is the interrogation time. During this process, the phase shifts experienced by the atoms are sensitive to inertial forces, resulting in an atomic interference fringe that provides acceleration and angular velocity information.

Over the past 30 years \cite{GeigerAVSQS2020, Tino21}, LPAIs have demonstrated excellent laboratory sensitivity \cite{Asenbaum20}, and serious efforts are underway to advance the platform for real-world applications \cite{BongsNatRevPhys2019} and multi-axis sensors \cite{Canuel06, Hardman16, WuOptica2017, Chen19}. Their applications range from fundamental physics \cite{Asenbaum20, RosiNature2014, CanuelSciRep2018, Zhan20, Kasevich22} and geophysics \cite{WuSciAdv2019, Holynski22} to civil engineering \cite{Bongs16} and navigation \cite{Naducci22}. Mobile LPAI prototypes have been demonstrated for land-based \cite{MenoretSciRep2018, WuSciAdv2019}, naval \cite{Bidel18}, and airborne \cite{GeigerNatCom2011, BarrettNatCom2016, Bidel20} use under micro-g to \SI{< 2}{g} environments, where g is the acceleration due to gravity. Furthermore, Bose-Einstein condensates have been tested under micro-g in sounding rockets \cite{BeckerNature2018} and an Earth-orbiting research laboratory \cite{Thompson20} with the aim of demonstrating LPAI functionality. However, significant physics and engineering challenges currently hinder the development of deployable cold-atom inertial sensors in dynamic environments. For example, it is difficult to simultaneously ensure that the sensor maintains high performance in extreme dynamics and that the form factor of the sensor head and the laser system remains compact. 

A trade-off arises in LPAI operation because the highest sensitivity is achieved with long-baseline atom interferometry (requiring a long $T$), but high dynamics can affect the relative movement between cold atoms and the interrogating laser beams (e.g., through on-axis acceleration, cross-axis acceleration, cross-axis rotation, and vibration) and therefore high data-rate LPAI operation is required to mitigate those effects (most effective with a short $T$). A short time-of-flight approach \cite{Biedermann12, Biedermann14, Bongs21} can produce an exceptional accelerometer that operates up to \SI{330}{\Hz}, opening up the possibility of LPAI operation in a high-g environment (e.g., \SIrange[range-units = single]{1.5}{9}{g} \cite{Nagy64, Schoenster71, SRUH15}) with modest sensitivity towards high accelerometer performance (e.g., bias stability < \SI{0.5}{\micro g}, scale factor < \SI{1}{ppm} \cite{Travagnin20}).

To address these challenges, we present progress on a multifaceted approach. Our sensor head has a compact optomechanical design that includes fixed optical components, custom Ti vacuum packages, and microfabricated grating chips. Our laser system architecture is based on a single seed laser and multiple single sideband modulators (SSBMs) with time-multiplexed frequency shifting, which is compatible with (and, thus, can be replaced by) integrated photonic components. Using the sensor head and the laser system, we demonstrate a grating magneto-optical trap (GMOT) \cite{Arnold10, Hinds13, Lee13, Squires17, Arnold17, BarkerPRAp2019, SitaramRSI2020, McGeheeNJP2021, Seo:21} with a single laser-cooling beam, a high data-rate sub-Doppler-cooled GMOT (\SI{15}{\micro\K}, \SI{20}{Hz}), Ramsey interferometry with the GMOT, and an initial demonstration of a cold-atom accelerometer based on the GMOT ($\Delta g / g = \num{2.0e-06}$, $T = \SIrange[range-units=single, range-phrase=-]{0}{4.5}{\ms}$, \SI{10}{Hz}). Our LPAI sensor head is designed to perform counter-propagating Doppler-sensitive Raman without a retroreflecting mirror. This is in contrast to most LPAI systems \cite{MenoretSciRep2018, WuSciAdv2019, Bidel18, GeigerNatCom2011, Bidel20}, which rely on a retroreflecting mirror to maintain the phase between two Raman beams, an approach that makes it difficult to distinguish oppositely diffracted atoms \cite{PerrinPRA2019}. Eliminating the retroreflecting mirror removes the degeneracy of low-velocity atoms present in retroreflecting Raman LPAIs \cite{PerrinPRA2019}, enabling high data-rate operation.

To ensure that the LPAI can operate in high-dynamic conditions, the laser system must be designed to withstand such conditions since it is subjected to similar dynamics. Substantial efforts have been made towards the development of compact and robust laser systems \cite{LevequeAPB2014, SchkolnikAPB2016, CaldaniEPJD2019, SabulskySciRep2020, FryeEPJQT2021}, including recent demonstrations with SSBMs \cite{ZhuOE2018, RammelooJOSAB2020} and single seed lasers \cite{WuOptica2017, ChiowAPB2018, Fang:18, MacraeOL2021}. Many of these systems are based on discrete components with fiber-to-fiber connections or free-space optical path connections (with optomechanical alignment mounts), which limit their ability to withstand high dynamics and severely limit manufacturing scalability. Additive manufacturing \cite{Hackermuller21a, Hackermuller21b} may also provide pathways towards vacuum miniaturization and robust optics platforms.

To maximize manufacturability and performance with a small form-factor, we believe that microfabricated photonic integrated circuit (PIC) technology \cite{Kodigala19, Weigel:18, Boynton:20, Heck:11, Davis:20, Otterstrom:19, Moore:16, Siddiqui:19} can be a robust and reliable solution. In another effort by our team, we have developed a PIC laser architecture to use a telecom wavelength (\SI{1560}{nm}) for light modulation and optical amplification and use a frequency-doubled wavelength (\SI{780}{nm}) to address Rb atoms; we have performed demonstrations of critical PIC components \cite{Kodigala19}. In this paper, we demonstrate a laser system with similar architecture to the PIC-based concept using commercial-off-the-shelf (COTS) components. The combination of our PIC-compatible approach and our high data-rate GMOT will lead to reduced size, weight, and power (SWaP) of the LPAI sensor head, and, thus, represents an important step towards deployable cold-atom inertial sensors that are robust against vibration, shock, and radiation.

\section*{Results}

Our multifaceted approach for mobile LPAI operation includes a small form-factor sensor head with a microfabricated grating chip, custom vacuum package, and a compact optomechanical design with fixed optical components (\cref{fig:sensor_head}, \cref{fig:Ti_vacuum_package}, \cref{fig:sensor_head_optics}); a PIC-compatible laser system architecture (\cref{fig:laser_architecture}, \cref{fig:laser_system}); a high data-rate GMOT and Ramsey interferometry (\cref{fig:GMOT_data_rate}, \cref{fig:Coherence}); and a cold-atom accelerometer based on the GMOT (\cref{fig:AI_fringe}).

%%%%%%%%%% Compact Sensor Head %%%%%%%%%%

The compact sensor head is designed as a tethered system with a remote electronics and laser system connected via cables and optical fibers (see \cref{fig:sensor_head}(a)). The optomechanical design of the small form-factor sensor head supports a miniaturized \ce{Ti} vacuum package (including an in-vacuum-mounted microfabricated grating chip) and provides optical access for laser cooling and trapping, Doppler-sensitive Raman transitions, and the fluorescence detection of the atoms. To maximize both manufacturability and efficiency, we chose a GMOT \cite{Arnold10, Hinds13, Lee13, Squires17, Arnold17, BarkerPRAp2019, SitaramRSI2020, McGeheeNJP2021, Seo:21} in the production of cold \Rb{87} atoms for the LPAI because its simple, streamlined design uses a single incoming cooling beam to ensure superior optical access and eliminate unnecessary mechanical degrees of freedom (see \cref{fig:sensor_head}(b)).

\subsection*{Microfabricated Grating Chip for a GMOT}

For the compact LPAI sensor head, we designed a microfabricated reflective grating chip to form the GMOT \cite{Arnold10, Hinds13, Lee13, Squires17, Arnold17, BarkerPRAp2019, SitaramRSI2020, McGeheeNJP2021, Seo:21}, which uses a single incoming laser-cooling beam and multiple first-order diffracted beams.
The reflective grating chip is based on a triangular geometry with three 1D binary grating sections that generate a tetrahedral GMOT.
Although the diffraction efficiency of binary gratings is lower than that of blazed and multilevel gratings, they are simpler to fabricate.
First, we designed \cite{gdcalc} a binary grating for the rubidium D2 transition at \SI{780.24}{\nm} \cite{Steck87RbNumbers} to meet a variety of design parameters, including pitch, duty cycle, depth, polarization, and reflective surface material; we simulated first-order and zeroth-order diffraction efficiencies of a binary grating (\SI{\sim 50}{\percent} duty-cycle and \SI{195}{\nm} depth) for $s$-polarized light (\SI{37.88}{\percent}, \SI{14.29}{\percent}), $p$-polarized light (\SI{47.75}{\percent}, \SI{0.92}{\percent}), and circularly polarized light (\SI{40.81}{\percent}, \SI{7.61}{\percent}).
The first order diffraction angle was \SI{40.56}{\degree} from the incoming beam, which is determined by $\theta_{\rm{m}} = \sin^{-1}(\frac{\text{m} \lambda}{\text{d}})$ for diffraction order $\text{m} = 1$, pitch $\text{d} = \SI{1.2}{\um}$, and wavelength $\lambda = \SI{780.24}{\nm}$. The fabrication process of the grating chip is shown in \cref{fig:grating_chip_fabrication}(b) (see Methods). We made the grating pattern in a circular shape (\SI{17.5}{\mm} diameter) to match the fused-silica viewport (\SI{\sim 19}{\mm} diameter) on the vacuum package. In our GMOT, the estimated capture volume was \SI{\approx 0.17}{\cm\cubed}.
We also diced the grating chip in a hexagonal shape and used the non-grating flat edges for the retainer clips (see \cref{fig:Ti_vacuum_package}(b) and \cref{fig:grating_chip_fabrication}(a)).

\subsection*{Custom Vacuum Package with the Grating Chip}

\begin{figure}[b!]
    \centering
	\includegraphics[width=1\columnwidth]{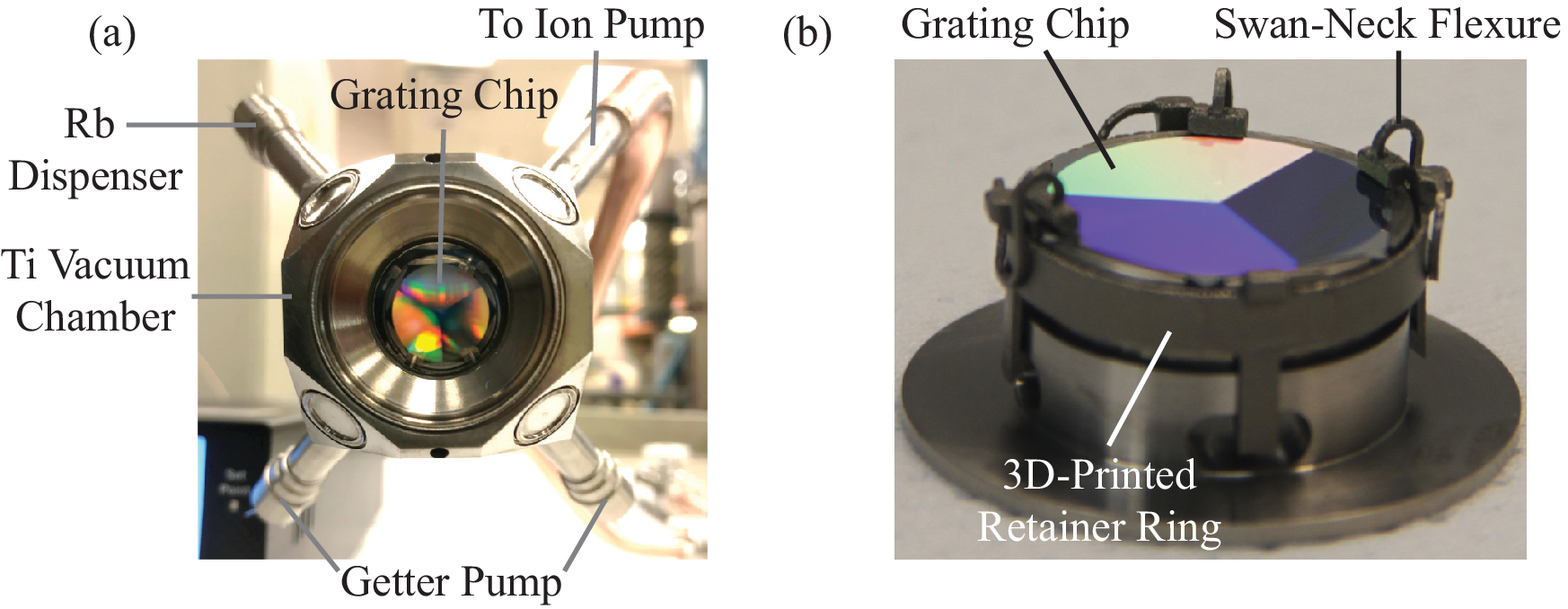}
	\caption{
			Images of the custom \ce{Ti} vacuum package and the in-vacuum mounted grating chip.
			(a) View of the vacuum package looking at the grating chip through the GMOT cooling beam entry window (\SI{\approx 19}{\mm} clear aperture).
			The external dimensions of a truncated Ti cube are approximately 40.64\,$\times$\,40.64\,$\times$\,40.64\,$\rm{mm}^3$.
			Also visible are \ce{Ti} tubes which house the rubidium dispensers, getters, and a brazed connection to the \ce{Cu} exhaust port.
			(b) Close-up view of the in-vacuum grating chip and mount prior to being laser-welded to the vacuum package.
			The grating chip is held in place with 3D-printed swan-neck flexure mounts.
		}
	\label{fig:Ti_vacuum_package}
\end{figure}

The ultra-high vacuum (UHV) package is the centerpiece around which our sensor head was built (\cref{fig:Ti_vacuum_package}(a)).
Briefly, we constructed the vacuum package from Grade-2 titanium (\ce{Ti}) with five anti-reflection coated fused silica viewports, non-evaporable getters (SAES St172/HI/7.5-7), and alkali-metal dispensers (SAES Rb/NF/3.4/12FT10+10) housed inside \ce{Ti} tubing laser-welded to the body.
Compared to steel, \ce{Ti} is nonmagnetic, has a lower thermal expansion coefficient, and has a much lower hydrogen outgassing rate.
We also brazed a copper exhaust port to the body for evacuation and connected it to an ion pump to maintain the vacuum conditions.

The GMOT grating chip was mounted internally and held with a sintered metal 3D-printed, \ce{Ti}-alloy (ASTM \ce{Ti-6Al-4V}) retainer that has four swan-neck flexure mounts, as shown in \cref{fig:Ti_vacuum_package}(b).
The 3D-printed parts were vacuum baked at \SI{450}{\degreeCelsius} prior to grating chip installation.
The flexure mounts provided optical access down to the surface of the grating chip for the Raman beams traveling parallel to the surface and avoided the use of epoxies.

The design of this vacuum package is similar to the passively pumped vacuum package \cite{Little2021passively} that has now demonstrated more than two years of magneto-optical trap (MOT) operation without active pumping.
Passive pumping eliminates the need for an ion pump, reducing system complexity and SWaP.
The passively pumped design in our previous work \cite{Little2021passively} does not include the grating chip and uses C-cut sapphire windows (for reduced birefringence and low-helium permeation), but future designs will include the grating chip.

\subsection*{Compact LPAI Sensor Head with Fixed Optical Components}

\begin{figure}[b!]
    \centering
	\includegraphics[width=0.75\columnwidth]{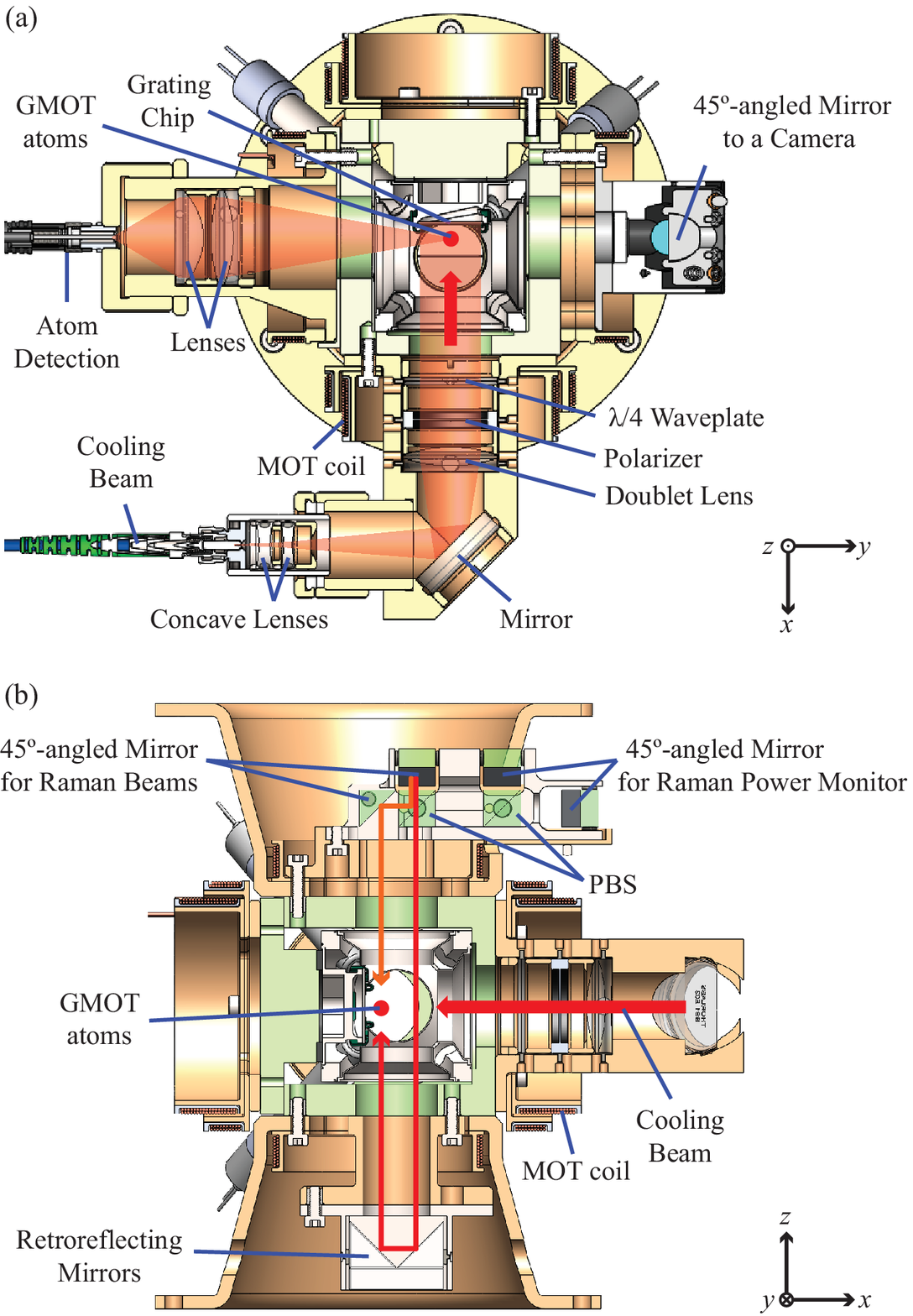}
	\caption{
			Cross-sectional renderings of the LPAI sensor head.
			(a) Horizontal cross-section showing the cooling-beam and atom-detection channels with fixed optical components.
			The cooling-channel light is delivered to the sensor head via a polarization maintaining (PM) fiber from which a large collimated Gaussian beam ($\rm{D}_{1/e^2} \approx \SI{28}{\mm}$) is used for cooling.
			The beam is truncated to \SI{\approx 19}{\mm}-diameter through the fused silica viewport in the compact LPAI sensor head.
			The light then passes through a polarizer and a $\lambda/4$ waveplate before illuminating the grating chip.
			The GMOT atoms (solid red circle) form \SI{\approx 3.5}{\mm} from the grating surface.
			The atom-detection channel was designed to measure atomic fluorescence through a multimode-fiber-coupled avalanche photodiode (APD) module.
			(b) Vertical cross-section of the sensor head showing the designed beam paths for Doppler-sensitive Raman.
			Cross-linearly-polarized Raman beams are launched from the same PM fiber and the two components are split by a polarizing beam splitter (PBS).
			Fixed optics route the Raman beams to the GMOT atoms (solid red circle) with opposite directions.
			Note: The data of \cref{fig:Coherence}(b) and \cref{fig:AI_fringe} were measured with free-space Raman beam optics.			
			}
	\label{fig:sensor_head_optics}
\end{figure}

To achieve deployable cold-atom inertial sensors, we sought to eliminate most of the optomechanical alignment components in the compact sensor head, which was designed to operate in high-vibration and harsh-temperature conditions.
In particular, the simplicity of a GMOT approach enables the elimination of unnecessary mechanical degrees of freedom and achieves a compact and reliable optomechanical design.
In addition, the time-multiplexed frequency shifting of a single seed laser simplifies the required optical channels for the tethered sensor head.
  
The optomechanical structure supports the vacuum package and provides the necessary optical routing for the GMOT, atom detection, and Raman beams, while making the sensor head more robust against vibration.
As shown in \cref{fig:sensor_head_optics}, the main structure surrounding the vacuum package primarily consists of a combination of FR4 fiberglass composite (green) and \SI{30}{\percent} glass-filled PEEK (brown). These non-conductive materials near the coils reduce the effects of eddy currents. We mechanically held the optical components while curing the epoxy (3M Scotch-Weld EC 2216, certified for use in aircraft and aerospace applications).

In \cref{fig:sensor_head_optics}(a), the horizontal cross-section view of the compact LPAI sensor head shows the fixed optical components for cooling-beam and atom-detection channels for the GMOT. 
When the grating chip is in operation (see \cref{fig:sensor_head}(b) and \cref{fig:Ti_vacuum_package}(b)), a single incoming laser-cooling beam and three first-order diffracted beams form a tetrahedral GMOT with a quadrupole magnetic field created by anti-Helmholtz coils.
All four cooling beams and the magnetic field zero point have to be aligned to the same point in 3D space.
We achieved the final alignment of the cooling beam to the grating using a 6-axis piezo stage that actively aligns the fiber-port-concave-lens assembly while monitoring the atomic cloud fluorescence via a CMOS camera (Imaging Source DMM 37UX265-ML) attached to the sensor head.
After that, all the optics were bonded in place.
For the steady-state GMOT atoms, the diameter of the atomic clouds was \SI{\sim 0.8}{mm} perpendicular to the grating and \SI{\sim 1.7}{mm} parallel to the grating; the distance between the atomic clouds and the grating surface was \SI{\sim 3.5}{mm}.
The repump/detection beam was coupled through the same assembly (as the cooling beam) with the same polarization.

\cref{fig:sensor_head_optics}(b) shows our original design for counter-propagating Raman beams, and the vertical cross-section view of the compact LPAI sensor head, which includes the fixed optical components for cross-linearly-polarized Raman beams, Raman power monitor ports, and cooling beam channels.
In this design, the Raman light is brought to the sensor head with a polarization maintaining (PM) optical fiber where two Raman beams propagate along the fast and slow axes of the fiber.
The Raman fiber launch assembly collimates the light, and a \SI{45}{\degree}-angled mirror brings the light into the plane of \cref{fig:sensor_head_optics}(b).
After this, the two Raman beams are split by the first polarizing beam splitter (PBS) and become counter-propagating Doppler-sensitive Raman beams.
The downward and upward Raman beams are then applied to the atoms released from the GMOT, driving the state-dependent momentum kicks on the atoms for LPAI operation.
The Raman power monitor ports couple light into multimode fibers, so that the pulse amplitude of each Raman beam can be measured.
We initially performed the alignment without the GMOT atoms using high-precision 6-axis piezo stages such that the Raman beams intersected at their designed position.
However, in practice, we found a mismatch between the optimized GMOT position and the designed Raman beam position.
We note that the data shown below (\cref{fig:Coherence}(b) and \cref{fig:AI_fringe}) was taken by substituting the fixed Raman optics with conventional free-space optics in order to optimize LPAI performance.
This mismatch can be suppressed in subsequent versions of the sensor head and further reduced with active alignment or flat-top Raman beams \cite{MielecAPL2018} to improve spatial-overlap homogeneity.

Our detection collects the atomic fluorescence of the GMOT atoms from a spherical region \SI{\sim 3}{\mm} in diameter, where the atomic cloud exists during the LPAI operation. It consists of two focusing elements and a multimode-fiber-coupled avalanche photodiode (APD), as shown in \cref{fig:sensor_head_optics}(a). The optics were centered in their housing using shims, secured into the housing with a threaded retaining ring to maximize the atomic fluorescence, then bonded into place.

%%%%%%%%%% PIC Laser System %%%%%%%%%%

\subsection*{Photonic-Integrated-Circuit-Compatible Laser System}

\begin{figure}[b!]
	\centering
	\includegraphics[width=1\columnwidth]{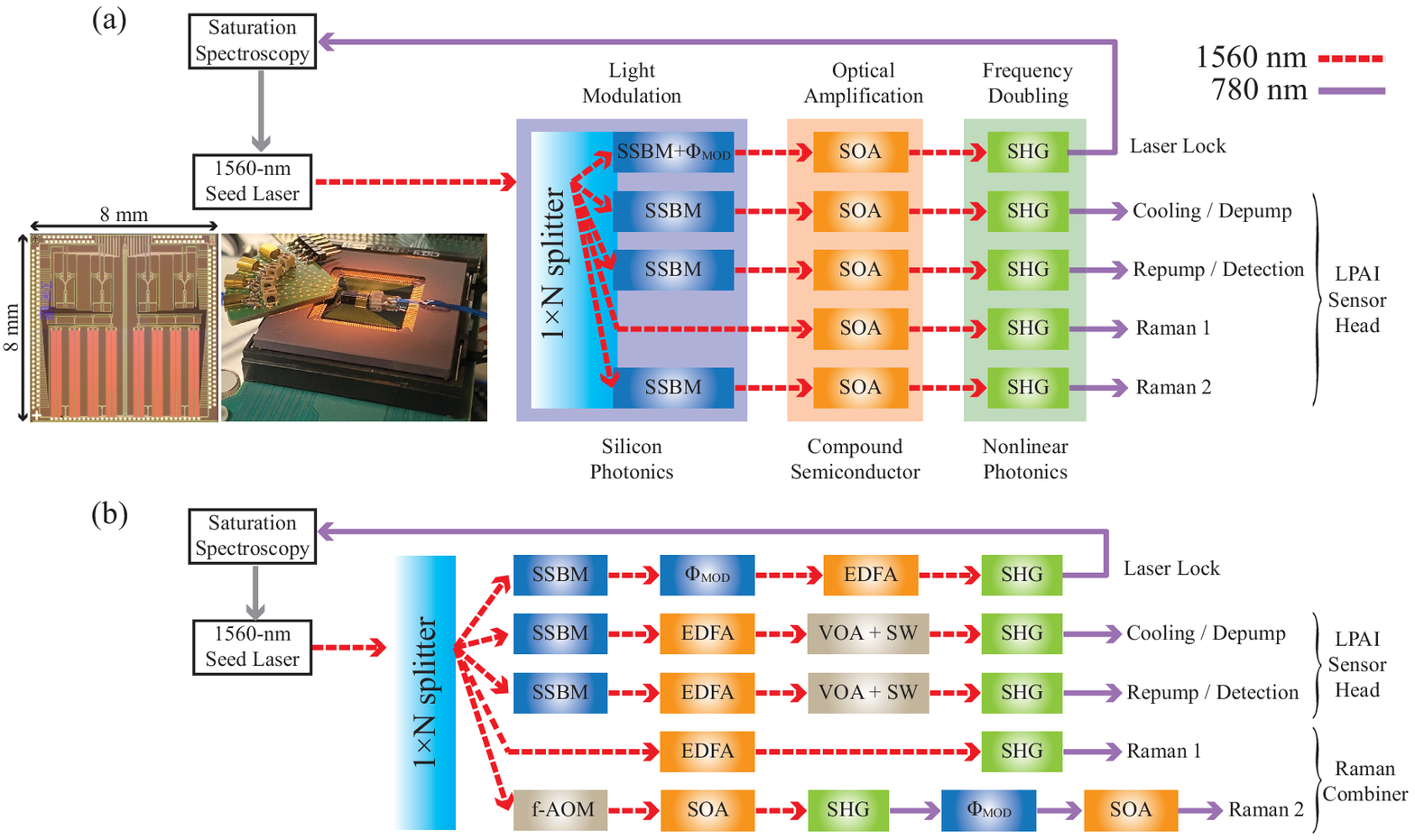}
	\caption{The laser architecture of the LPAI with a single seed laser and time-multiplexed frequency shifting.
			(a) PIC-based laser system composed of silicon photonics for light modulation, compound semiconductor for optical amplification, and nonlinear photonics for frequency doubling towards a co-packaged laser system and a single hybrid-integrated PIC laser system. (Inset-Left) This image is a four-channel silicon photonic SSBM chip with dual-parallel Mach-Zehnder modulators, simultaneously developed at Sandia \cite{Kodigala19}. This PIC chip ($8 \times 8$ $\textrm{mm}^2$) includes 17 silicon photonic phase modulators (4 for each SSBM), variable optical attenuators, thermo-optic phase shifters, optical filters, and photo detectors. (Inset-Right) This image is a fully packaged silicon photonic SSBM, which has been used to generate the Raman beams from a laser in an LPAI demo \cite{Kodigala22}.
			SSBM, single sideband modulator; $\Phi_{\rm{MOD}}$, phase modulator; SOA, semiconductor optical amplifier; SHG, second harmonic generator.
			Each channel in the silicon photonics includes a VOA to control the amplitude of the light, and an SOA can function as an optical switch. 
			(b) PIC-compatible laser system based on discrete components used for the data shown in this paper. f-AOM, fiber-coupled acousto-optic modulator; EDFA, erbium doped fiber amplifier; VOA, variable optical attenuator; SW, optical switch. For Raman 2, we used a f-AOM at 1560 nm and a $\Phi_{\rm{MOD}}$ at 780 nm, which can be replaced with a SSBM at 1560 nm similar to the PIC-based laser system. The f-AOM is used for the phase lock by continuously monitoring the beat note between the two Raman beams.
			\cref{fig:raman_combiner} shows the beat note measurement and the switching of the Raman beams.
		}
	\label{fig:laser_architecture}
\end{figure}

\begin{figure}[b!]
	\centering
	\includegraphics[width=1\columnwidth]{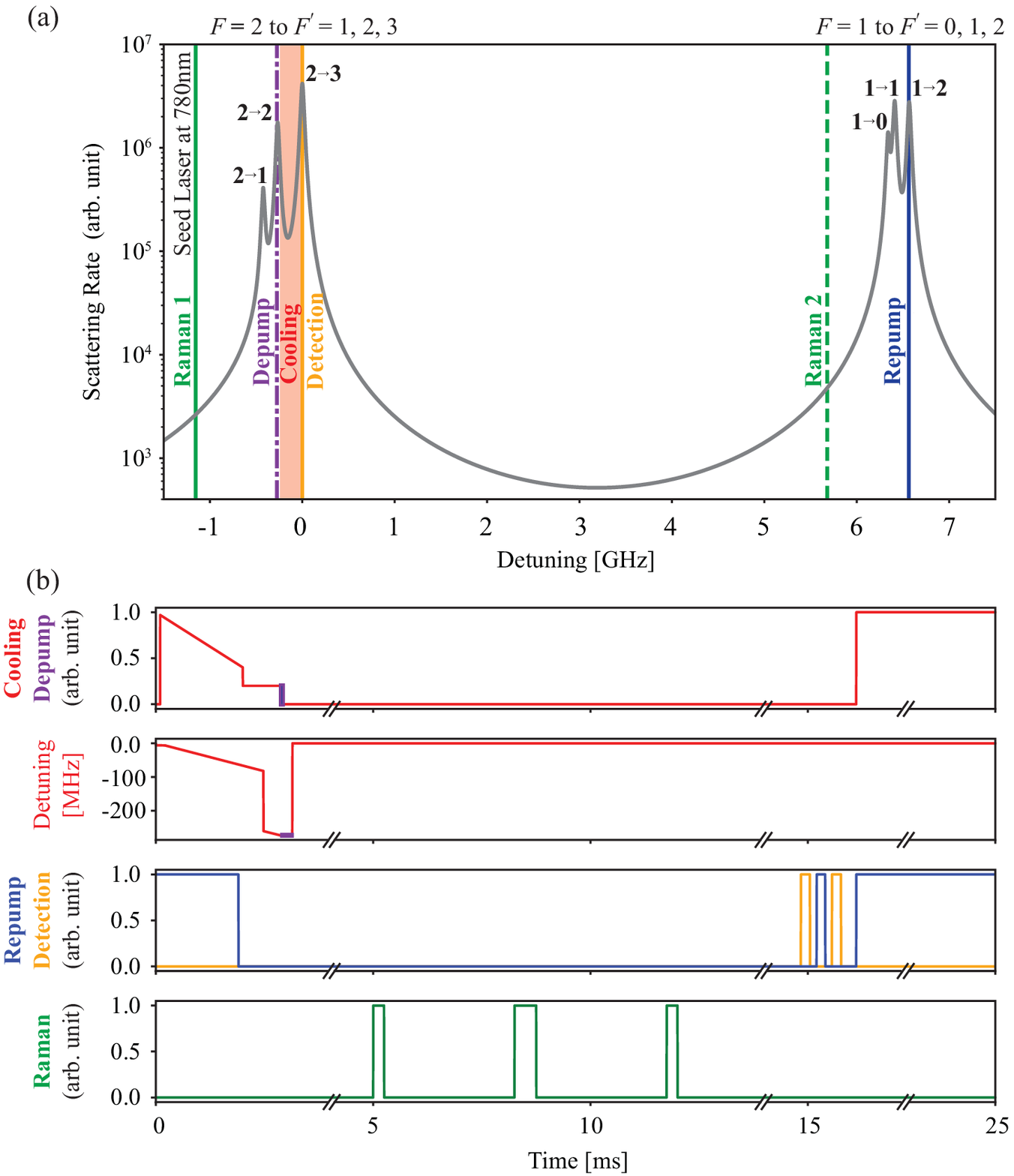}
	\caption{
			Optical frequency and timing diagram for the PIC-compatible laser system.
			(a) Optical transitions of \Rb{87} atoms (D2 transition) \cite{Steck87RbNumbers} for LPAI operation. 
			Cooling, Detection, Depump, and Raman 1 beams use $\ket{F = 2}$ to $\ket{F' = 1, 2, 3}$ transitions (Left peaks), and Repump and Raman 2 beams use $\ket{F = 1}$ to $\ket{F' = 0, 1, 2}$ transitions (Right peaks). The \SI{1560}{nm} seed laser is frequency-doubled to \SI{780}{nm} and locked to $\ket{F = 2}$ to $\ket{F' = 2, 3}$ crossover transition with \SI{\sim 1.15}{GHz} red-detuning from the $\ket{F = 2} \rightarrow \ket{F' = 3}$ resonance, which is also used for Raman 1.
			(b) Diagram of the timing sequence for LPAI operation highlighting important dynamic changes in the laser channels. The pulse width of Repump/Detection/Raman is not to scale.
		}
	\label{fig:laser_system}
\end{figure}

The development of a compact high-performance laser system is essential to reduce design complexity and improve reliability in a dynamic environment.
The long-term goal is to achieve extreme miniaturization of the laser system by integrating nearly all of the required optical components onto a single PIC.
Over the last decade, silicon-based PICs operating in the telecom C-band have matured substantially, providing the components required for optical telecommunication.
In addition, the heterogeneous integration of PIC components is ongoing; for example, silicon photonics can be integrated with compound semiconductors to form lasers and optical amplifiers \cite{Heck:11,Davis:20}, nonlinear material (e.g., thin-film lithium niobate) to create electro-optic modulators \cite{Weigel:18, Boynton:20}, and Brillouin-active membrane to form Brillouin lasers and amplifiers \cite{Otterstrom:19}. In addition, nonlinear photonics (second harmonic generation, SHG) has been demonstrated on a silicon-photonics-compatible platform with lithium niobate resonators \cite{Moore:16} and aluminum nitride waveguides \cite{Siddiqui:19}.

These ongoing efforts would bring modulators, optical amplifiers, frequency doublers, and even laser sources onto the PIC to achieve a fully integrated laser solution for an LPAI sensor. Using a PIC, the optical frequency can be manipulated at \SI{1560}{\nm} and then frequency-doubled to \SI{780}{\nm} (the \ce{Rb} D2 line).
Because chip-based technologies are highly scalable, moving from a single-axis accelerometer to a three-axis accelerometer or a complete LPAI-based inertial measurement unit can be readily envisioned using PIC technology.
Here, we outlined a potential PIC-based concept (for a PIC laser architecture) and demonstrated a "PIC-compatible" laser system, showing that the architecture is functional for an LPAI using discrete COTS components.

As shown in \cref{fig:laser_architecture}(a), our PIC-based laser system (based on a 1560-to-\SI{780}{nm} approach), aims to realize a mass-producible, chip-scale laser system with integrated photonics for laser cooling and trapping and atom interferometry. This laser system has three major functions: light modulation (silicon photonics), optical amplification (compound semiconductor), and frequency doubling (nonlinear photonics). A silicon PIC chip achieves light modulation using multichannel SSBMs \cite{Kodigala19}, where a $1 \times N$ silicon nitride splitter divides an off-chip seed laser (\SI{1560}{nm}) into multiple channels. Within each channel, a SSBM is used to frequency offset the light to the required frequency for the particular function of the channel. The SSBM allows the light frequency to be rapidly switched (time-multiplexed frequency shifting), allowing a single channel to perform multiple functions, such as repump light for the MOT and detection light for state read out (see \cref{fig:laser_architecture} and \cref{fig:laser_system}). A 1560-nm wavelength silicon PIC chip \cite{Kodigala19} includes frequency/phase modulators \cite{DeRose12}, thermo-optic phase shifters, variable optical attenuators (VOAs), optical filters, and photo-detectors (to monitor the performance of the other components). After passing through the SSBM, the light goes into a \SI{1560}{nm} high-power semiconductor optical amplifier (SOA) that produces up to \SI{500}{mW}, and then passes through a high-efficiency frequency doubler to convert the light to \SI{780}{nm}.

The architecture shown in \cref{fig:laser_architecture}(a) was chosen because the essential components can be realized (or are close to being realized) with existing PIC technologies. By developing these building blocks, our team has successfully implemented a four-channel SSBM on a silicon PIC (\cref{fig:laser_architecture}(a) inset) and performed an initial demonstration with the silicon PIC to generate the Raman beams in an LPAI \cite{Kodigala22}. Our team is currently developing a high-power multi-stage SOA with compound semiconductor and a nonlinear photonics frequency doubler. This SOA could also operate as a high-speed (sub-nanosecond) optical switch\cite{Ju:05} with a high extinction ratio. Using a lithium niobate platform with shallow etched waveguides\cite{ZhaoOE20} for the frequency doubler, our team demonstrated \SI{939}{\percent\per\W} SHG conversion efficiency. While \cref{fig:laser_architecture}(a) shows separate PICs for the silicon photonics, compound semiconductor, and nonlinear photonics, we envision a future PIC where all required optical components are hybrid-integrated into a single PIC.

The PIC-compatible laser system is shown in \cref{fig:laser_architecture}(b), where essential functions of the PIC architecture are replaced with fiber-connected COTS components.
A single \SI{1560}{\nm} fiber seed laser with \SI{\leq 3}{\kHz} linewidth (NP Photonics FLM-150-3-1560.49-1-S-A) splits into five channels: (1) Lock, (2) Cooling, (3) Repump/Detection, (4) Raman 1, and (5) Raman 2. Channels 1, 2, 3 share a similar architecture, frequency-shifting the light from the seed laser using the SSBM \cite{ZhuOE2018, RammelooJOSAB2020} (IQ modulator, Eospace IQ-0D6S-25-PFA-PFA-LV).
The Lock channel has an additional phase modulator ($\Phi_{\rm{MOD}}$, Eospace PM-0S5-10-PFA-PFA-UL) placed after the SSBM to stabilize the frequency of the seed laser to a \Rb{87} reference ($\ket{F = 2} \rightarrow \ket{F' = 2,3}$ crossover resonance in \Rb{87}) with frequency modulation saturation spectroscopy.
The \SI{1560}{\nm} light is then amplified by an erbium-doped fiber amplifier (EDFA, IPG Photonics EAR-2K-C-LP-SF) and passes through a high-power electro-optic VOA (Agiltron NHOA-325115333) and a high-power electro-optic optical switch (SW, Agiltron NHSW-115115333) before frequency doubling to \SI{780}{\nm} with the fiber-SHG modules (SHG, AdvR RSH-T0780-P15P78AL0).

The Cooling channel provides the cooling light on the $\ket{F = 2} \rightarrow \ket{F' = 3}$ transition with an intensity of about \SI{1.1}{\mW\per\cm\squared} and a detuning of $-1\,\Gamma$ for steady-state GMOT operation  (see \cref{fig:laser_system} (a)), where $\Gamma/ 2\pi = $ \SI{6.065}{MHz} for \Rb{87} D2 transition.
After the GMOT is loaded, sub-Doppler cooling is accomplished with a frequency ramp from  $-1\,\Gamma$ to $-13.5\,\Gamma$ and a VOA-controlled intensity ramp (\cref{fig:laser_system}(b)). We determined a \SI{15}{\micro\K} temperature of sub-Doppler cooled GMOT atoms by evaluating the width of the Doppler-sensitive Raman resonance. After sub-Doppler cooling, the repump light is extinguished and the Cooling channel jumps to the $\ket{F = 2} \rightarrow \ket{F' = 2}$ transition to depump the atoms into the $\ket{F = 1}$ hyperfine ground state  (\cref{fig:laser_system}(b)).

For fluorescence detection, the Repump/Detection channel is first resonant with the $\ket{F = 2} \rightarrow \ket{F' = 3}$ transition to detect atoms in the $\ket{F = 2}$ state  (see \cref{fig:laser_system}(a)).
It then jumps to $\ket{F = 1} \rightarrow \ket{F' = 2}$ to repump all atoms into the $\ket{F = 2}$ state before jumping back to the $\ket{F = 2} \rightarrow \ket{F' = 3}$ transition to detect all of the atoms (\cref{fig:laser_system} (b)).
For our experiment, we coupled the Repump/Detection channel to the same fiber as the Cooling channel with the same polarization.

The cross-linearly-polarized Raman beams control the internal states of the atoms, which are composed of two light frequency components (Raman 1 and Raman 2) using stimulated Raman transitions (\cref{fig:AI_fringe}(a)). In particular, Doppler-sensitive Raman beams provide the state-dependent momentum kicks required to construct atom interferometry in the space-time trajectory and measure inertial forces on the atoms.
We derived both Raman frequencies from the same \SI{1560}{\nm} seed laser, as shown in \cref{fig:laser_architecture}(b).
Raman 1 is the frequency-doubled seed laser (\SI{780}{\nm}) that is \SI{\sim 1.15}{GHz} red-detuned from the $\ket{F = 2} \rightarrow \ket{F' = 3}$ resonance.
Raman 2 is frequency-offset via a fiber-coupled \SI{1560}{\nm} acousto-optic modulator (AOM) operating at \SI{-150}{MHz}.
The frequency-offset light is then amplified by a \SI{1560}{\nm} SOA and frequency-doubled to \SI{780}{\nm}.
After that, Raman 2 is phase-modulated by a \SI{780}{\nm} phase modulator and then amplified by a \SI{780}{\nm} SOA. 
\cref{fig:raman_combiner} shows how the Raman beams are combined, with \SI{10}{\percent} of the light used for phase locking (see Methods). In the results shown below, Raman spectroscopy was applied to Ramsey interferometry with co-propagating Doppler-free Raman beams (\cref{fig:Coherence}(b)) and atom interferometry with counter-propagating Doppler-sensitive Raman beams (\cref{fig:AI_fringe}(c)).\\

%%%%%%%%%% Experimental Demonstration %%%%%%%%%%

We developed the compact sensor head (\cref{fig:sensor_head}, \cref{fig:Ti_vacuum_package}, \cref{fig:sensor_head_optics}) and the PIC-compatible laser system (\cref{fig:laser_architecture}, \cref{fig:laser_system}) as a step towards deployable cold-atom inertial sensors.
Using these components, we demonstrated a high data-rate sub-Doppler-cooled GMOT with a single laser-cooling beam (\cref{fig:GMOT_data_rate}), validated atomic coherence with Ramsey interferometry (\cref{fig:Coherence}), and performed proof-of-concept LPAI operation and a measurement of gravity (\cref{fig:AI_fringe}).

\subsection*{High Data-Rate GMOT Operation}

\begin{figure}[t!]
    \centering
	\includegraphics[width=1\columnwidth]{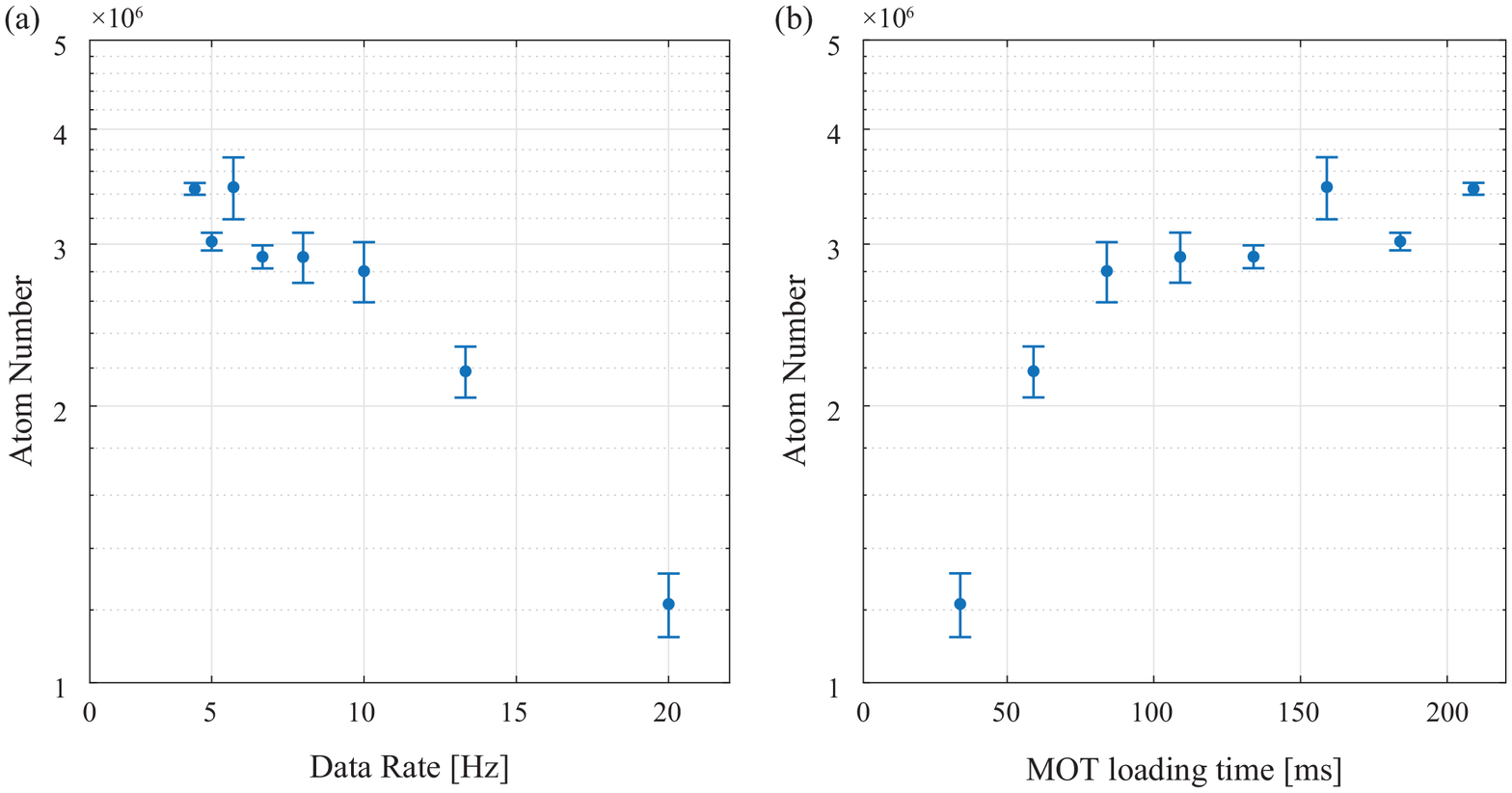}
	\caption{
			Measurement data rate and loading time of GMOT operation, including sub-Doppler cooling process.
			(a) Plot of atom number versus measurement data rate of GMOT operation.
			(b) Plot of atom number versus MOT loading time.
              Error bars are standard deviation.
		}
	\label{fig:GMOT_data_rate}
\end{figure}

A short time-of-flight approach enables high data-rate operation and enhances the bandwidth of LPAIs \cite{Biedermann12, Biedermann14} by mitigating the detrimental effects of sensor head movement during operation, which can ruggedize LPAIs for use in a dynamic environment with modest sensitivity. Efficient atom recapture is critical to maintaining a reasonably high atom number during high data-rate operation. In our system, high data-rate LPAI operation follows this sequence: (1) capture and cool atoms from background vapor with a GMOT, (2) perform sub-Doppler cooling, (3) prepare the atoms in the $\ket{F = 1}$ state, (4) release to free space, (5) perform physics experiments, (6) detect the atomic populations, and (7) recapture atoms for the next measurement cycle.

Here, we demonstrated proof-of-concept high data-rate GMOT operation by measuring the sub-Doppler cooled atom number vs. data rate (see \cref{fig:GMOT_data_rate}(a-b)).
For this measurement, we determined the atom number by the absorption of a probe beam resonant with the $\ket{F = 2} \rightarrow \ket{F' = 3}$ transition.
Within a measurement cycle, the MOT was loaded for a variable amount of time before the quadrupole magnetic field was turned off.
Sub-Doppler cooling and depumping to $\ket{F = 1}$ occurred over the next \SI{3}{\ms} and was followed by a \SI{13}{\ms} window for physics experiments and state detection.
At the end of the window, the cooling light and quadrupole field were turned back on to recapture atoms and restart MOT loading.
The data rate is the inverse of the total cycle time.
The number of sub-Doppler-cooled GMOT atoms decreased as the data rate increased, as shown in \cref{fig:GMOT_data_rate}(a); the same data is replotted in \cref{fig:GMOT_data_rate}(b), as the MOT loading time increased. 
At present, the data-rate measurement of GMOT operation is limited by the data rate of the control system and the low recapture efficiency, likely due to the low GMOT cooling intensity \cite{McGilliganOE2015}.

\subsection*{Ramsey Interferometry with a GMOT}

\begin{figure}[t!]
	\centering
	\includegraphics[width=1\columnwidth]{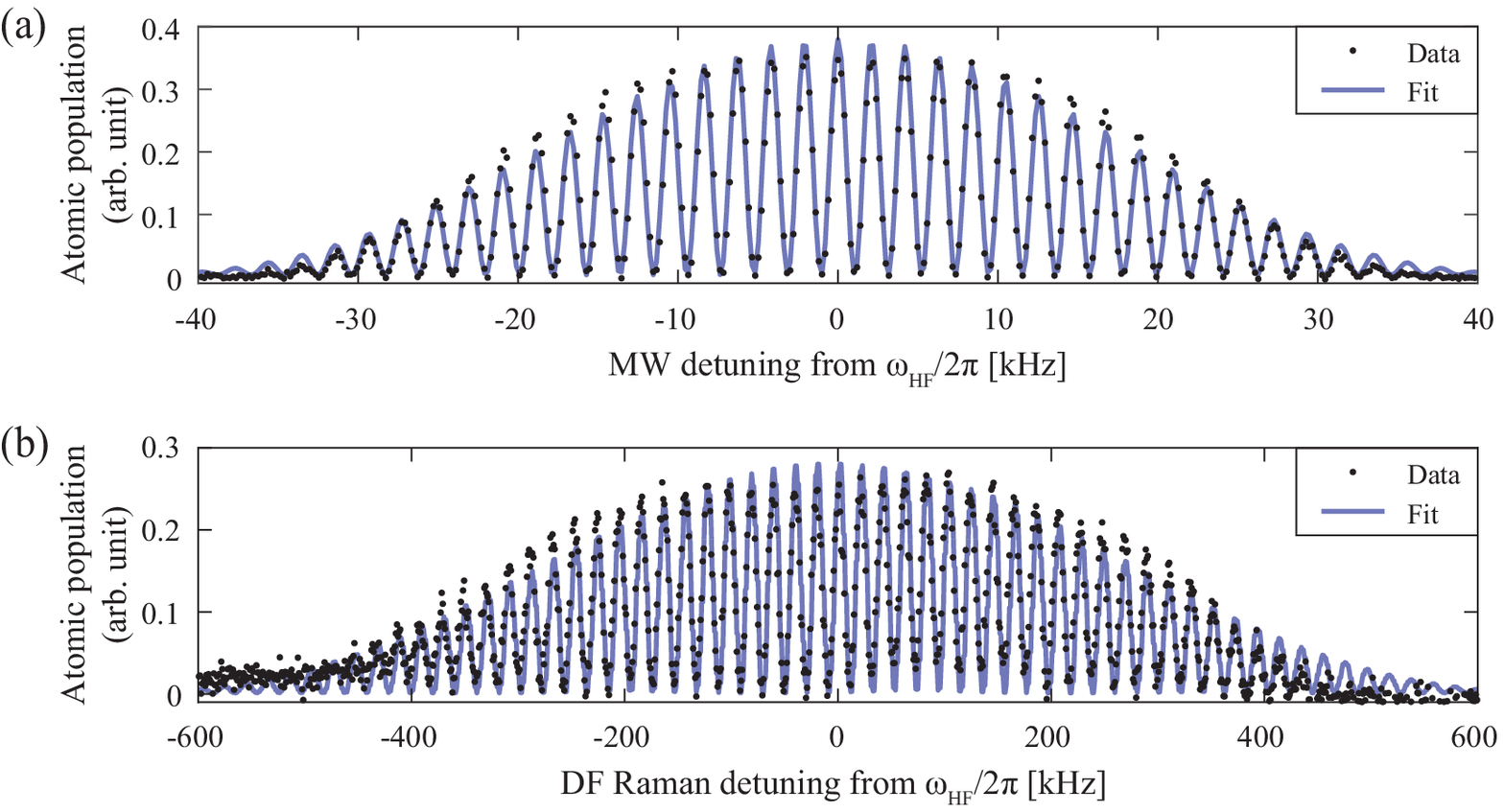}
	\caption{
			Ramsey interferometry ($\frac{\pi}{2} \rightarrow T \rightarrow \frac{\pi}{2}$) of sub-Doppler cooled GMOT atoms at the compact LPAI sensor head.        
			(a) The atomic fringe of Ramsey interferometry versus microwave detuning from atomic resonance ($\omega_{\rm{HF}}/2\pi$).
			The interrogation time $T$ is \SI{450}{\micro s}, resulting in a fringe spacing of $\delta\omega_{\rm{Ramsey}}/2\pi \simeq$ \SI{2.27}{\kHz}.
			The measurement data rate is \SI{13.33}{\Hz}.
			(b) The atomic fringe of Ramsey interferometry versus Doppler-free Raman detuning from atomic resonance ($\omega_{\rm{HF}}/2\pi$).
			The interrogation time $T$ is $\SI{48.08}{\micro s}$, resulting in a fringe spacing of $\delta\omega_{\rm{Ramsey}}/2\pi \simeq$ \SI{19.74}{\kHz}.
			The measurement data rate is \SI{10}{\Hz}. Both plots are based on a single shot-to-shot measurement.
			}
	\label{fig:Coherence}
\end{figure}

We examined atomic coherence to show that the system is not limited by decoherence due to background collisions/pressure and spatial-overlap inhomogeneity between cold atoms and Raman beams.
We observed Rabi oscillations and Ramsey fringes of the GMOT atoms using both a microwave field and co-propagating Doppler-free Raman beams that address the ``clock'' transition in the hyperfine ground-state manifold, i.e., $\ket{F = 1, m_F = 0}$ to $\ket{F = 2, m_F = 0}$.

In Rabi measurement with a microwave field, the Rabi frequency was \SI{10}{\kHz} ($\rm{T}_{2\pi} = $ \SI{100}{\micro s}), and the Rabi coherence time was $\tau_{\rm{1/e}} \simeq$ \SI{676}{\micro s}. 
In Rabi measurement with Doppler-free Raman beams, the Rabi frequency was \SI{130}{\kHz} ($\rm{T}_{2\pi} = \SI{7.68}{\micro s}$) and the Rabi coherence time was $\tau_{\rm{1/e}} \simeq$ \SI{8.3}{\micro s}.
The measured pulse times of the Rabi measurements were used for the Ramsey measurements.

For the Ramsey interferometry study ($\frac{\pi}{2} \rightarrow T \rightarrow \frac{\pi}{2}$), the frequency detuning of the two $\frac{\pi}{2}$-pulses was scanned for a fixed time $T$ to measure atomic interference fringes based on the two hyperfine ground states, as shown in \cref{fig:Coherence}.
In Ramsey measurement with a microwave field, the fringe spacing was \SI{2.27}{\kHz} from the data fitting, which was consistent with $1/T$ for $T = \SI{450}{\micro s}$  (see \cref{fig:Coherence}(a)).
In Ramsey measurement with Doppler-free Raman beams, the fringe spacing was \SI{19.74}{\kHz} from the data fitting, which also corresponds to $1/T$ for $T = \SI{48.08}{\micro s}$ (see \cref{fig:Coherence}(b)).

The improvement of spatial-overlap homogeneity between the GMOT atoms and the Raman beams becomes important to achieve the best performance in Ramsey interferometry and atom interferometry \cite{Butts11}. In the experiment, the current ratio of the diameter of the atomic clouds (parallel to the grating, \SI{\sim 1.7}{mm}) to the beam diameter of the Raman beams ($\rm{D_{1/e^2}} \simeq \SI{5.4}{\mm}$) was comparable to one third. The spatial-overlap homogeneity can be further enhanced with flat-top Raman beams \cite{MielecAPL2018} or large Raman beams \cite{MielecAPL2018}.

\subsection*{Light-Pulse Atom Interferometry with a GMOT}

\begin{figure}[t!]
	\centering
	\includegraphics[width=1\columnwidth]{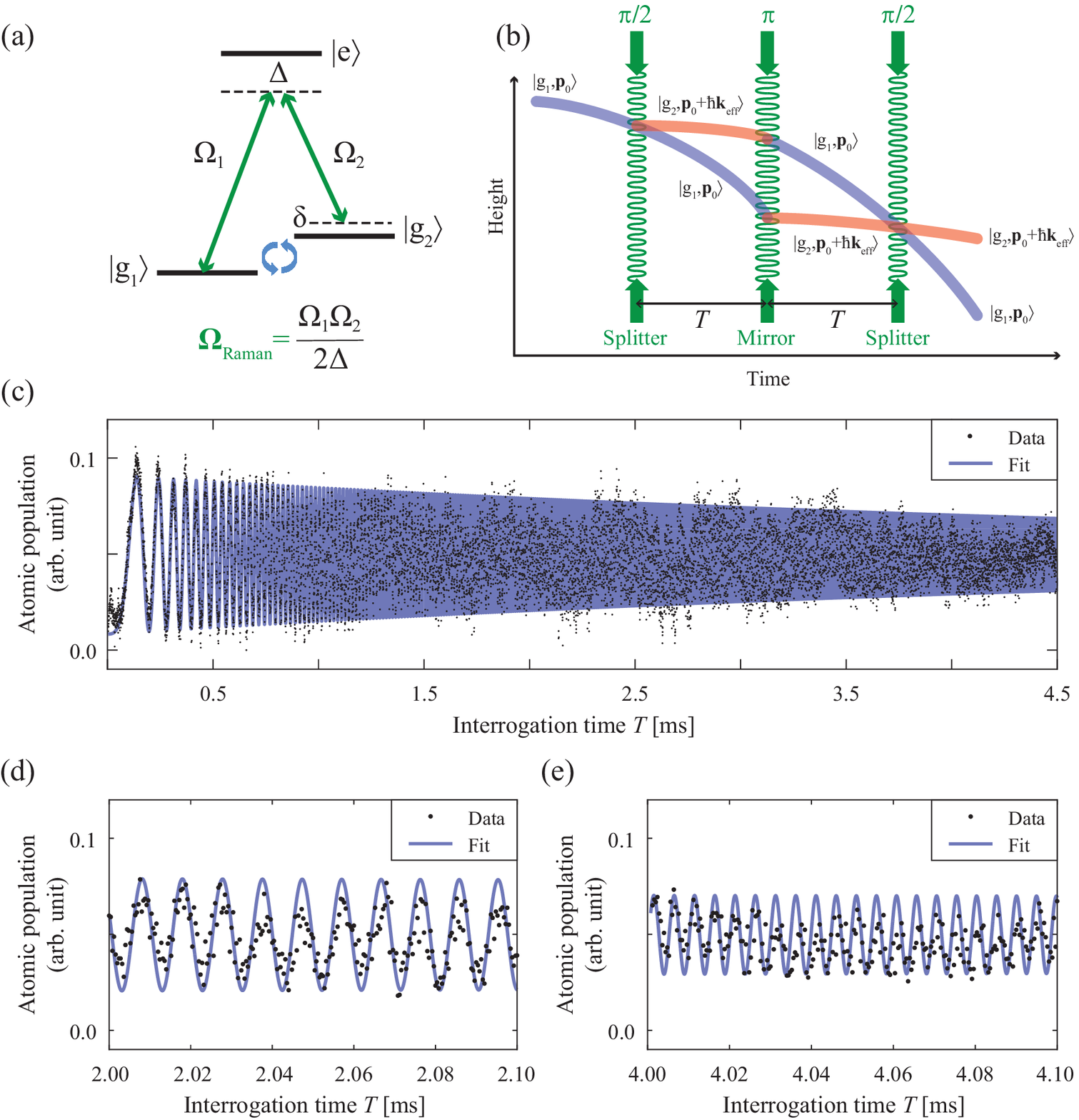}
	\caption{
			Atom interferometry of sub-Doppler cooled GMOT atoms in the compact LPAI sensor head.
			(a) Three-level atomic system for stimulated Raman transitions.
			$\ket{g_1} = \ket{F = 1, m_F = 0}$, ground state 1; $\ket{g_2} = \ket{F = 2, m_F = 0}$ ground state 2; $\ket{e}$, excited state; $\Delta$, single-photon detuning; $\delta$, two-photon detuning which depends on the Doppler shift; $\Omega_{\rm{Raman}}$, effective Rabi frequency of Raman beams; $\Omega_{1}$, single-photon Rabi frequency of $\ket{g_1} \rightarrow \ket{e}$ transition; $\Omega_{2}$, single-photon Rabi frequency of $\ket{g_2} \rightarrow \ket{e}$ transition.
			(b) Space-time trajectory of atomic wavepackets during LPAI.
			A three-light-pulse sequence, $\frac{\pi}{2} \rightarrow T \rightarrow \pi \rightarrow T \rightarrow \frac{\pi}{2}$, coherently addresses the ground states of the atoms and provides the state-dependent momentum kicks, $\hbar \textbf{k}_{\rm{eff}} \approx 2 \hbar \textbf{k}$, where $\textbf{p}_{0}$ is the initial atomic momentum; $T$, the free-evolution time between Raman pulses; $\textbf{k}_{\rm{eff}}$, the effective wavenumber of stimulated Raman transition; $\textbf{k}$, the wavenumber (${2\pi}/{\lambda}$) of a single Raman beam.
			(c) Scan of the interrogation time showing the LPAI chirped fringe for $T = \text{\SIrange{0.0}{4.5}{\ms}}$.
			The fringe chirping results from the Doppler shift of the cold atoms relative to the Raman beams as the atoms accelerate due to gravity.
			Each data point in the plot is an average of three data points and a slowly-varying offset was removed.
			The data is fit to a chirped sinusoid \cite{Peters98} $\cos\left(k_{\rm{eff}} \, g_{\rm{k}} \left( T^2 + \tau_{\rm{\pi}} (1 + \frac{2}{\pi}) T \right) + \phi_0 \right)$, where $\tau_{\rm{\pi}}$ is the $\pi$-pulse duration, $g_{\rm{k}}$ is gravity, and $\phi_0$ is an arbitrary phase factor.
			We measured $g = \SI{9.79316(2)}{\m\per\s\squared}$ with the statistical uncertainty of $\Delta g / g = \num{2.0e-06}$ without vibration isolation.
			Plots (d) and (e) provide a detailed view of the chirped fringe of (c) for $T = \text{\SIrange{2.0}{2.1}{\ms}}$ and $T = \text{\SIrange{4.0}{4.1}{\ms}}$, respectively.
			The measurement data rate is \SI{10}{\Hz}.
		}
	\label{fig:AI_fringe}
\end{figure}

We demonstrated proof-of-concept LPAI operation and gravity measurement using a GMOT, integrating all of our microfabricated grating chips, custom Ti vacuum packages, a compact-form-factor LPAI sensor head, and PIC-compatible laser systems with time-multiplexed frequency shifting.

In a gravimeter configuration, the LPAI accelerometer operates using stimulated Raman transitions (\cref{fig:AI_fringe}(a)), which coherently addresses the two hyperfine ground states of atoms, i.e., $\ket{F = 1, m_F = 0}$ and $\ket{F = 2, m_F = 0}$, and drives photon recoils on the atoms along the direction of gravity using counter-propagating Raman beams.
The three-light-pulse sequence of the LPAI coherently splits, redirects, and recombines atomic wavepackets for atomic interference in a space-time trajectory  (\cref{fig:AI_fringe}(b)).
The atomic interference fringe is sensitive to inertial forces that the atoms experience during the LPAI pulse sequence, and the phase shift provides gravitational acceleration information.

As discussed in \cref{fig:sensor_head_optics}, we optimized spatial overlap between the GMOT atoms and the Raman beams using free-space optomechanical components around the sensor head for Raman beams instead of the fully fixed optical components shown in \cref{fig:sensor_head_optics}(b).
The sensor head assembly was located on a countertop without any vibration isolation.
The LPAI accelerometer measures the gravitational acceleration by the phase shift of the atomic interference fringe, i.e., $\Delta \phi_{\text{acc}} = k_{\rm{eff}} \cdot a T^2$, when the sensing axis of Doppler-sensitive Raman beams is aligned to the gravity direction ($\textbf{a}$ = $\textbf{g}$).
The Raman quantization axis was defined along the Z-axis with a magnetic field of \SI{1.34}{G}, and the sensor head was located inside a single-layer magnetic shield.
Since the velocity of the atoms increased linearly due to $g$, the LPAI fringe became chirped as the interrogation time $T$ was increased.
Based on the fit of a chirped sinusoid \cite{Biedermann12}, we measured $g = \SI{9.79316(2)}{\m\per\s\squared}$ with LPAI operation up to $T = \SI{4.5}{\ms}$, as shown in \cref{fig:AI_fringe}(c-e).
The statistical uncertainty was $\Delta g / g = \num{2.0e-06}$ from the fitting the data for $T = \SIrange[range-units=single, range-phrase=-]{0.0}{4.5}{\ms}$.

\section*{Discussion}

We demonstrated a compact-form-factor high-data-rate atom interferometer with a PIC-compatible laser system architecture. One of our primary goals was to enable operation of the LPAI in high-dynamic environments (environments with large jerk), so our LPAI subsystems were designed with robustness in mind. First, the compact cold-atom sensor head used fixed optical components, a custom vacuum package, and a microfabricated grating chip for the GMOT \textendash\, all designed to enable ruggedization and miniaturization of the system. Using rigid materials for the miniaturized fixed-optics assembly is essential for decreasing the relative movement of components under changes in acceleration. Random vibration analysis of the compact sensor head showed good stability for the GMOT, but improvements are necessary for the Raman beam path for large changes in acceleration.  Second, the high data-rate operation of the LPAI using the sub-Doppler cooled atoms from the GMOT can enable operation through large dynamics in a number of ways. For example, decreasing the drop time reduces the sensitivity, but also keeps the size of the system small, minimizes the relative movement between cold atoms and the Raman beams (reducing the effects of cross-axis acceleration), and allows recapture of the atoms. While not discussed in this paper, a feedforward scheme can also enable operation over a large dynamic range \cite{Soh20}; in this scheme, conventional sensors measure inertial forces at a rate much larger than 1/T, and these signals are processed to ``feedforward'' to the Raman lasers (primarily the phase, but also the amplitude and frequency) to lock the LPAI to the side of an LPAI fringe. The feedforward scheme addresses mainly on-axis acceleration but can also mitigate cross-axis acceleration, cross-axis rotation, and vibration-induced noise on an LPAI fringe. Third, a PIC laser architecture is the ultimate in miniaturization for the laser system, and typically, the smaller the system size, the more dimensionally stable it is, allowing more tolerance to vibration. As part of our overall LPAI effort, we simultaneously developed the PIC-based laser system components (Fig. 5(a) inset), and validated our conceptual approach by using the PIC-compatible laser system for the LPAI demonstration. 

Our results show the potential for compact, multi-axis, and deployable cold-atom inertial sensors with microfabricated grating chips, although continued effort will be required to realize a fully integrated LPAI system for dynamic environments. Near-term improvements to our sensor head include a redesign of the Raman opto-mechanics to optimize overlap with the atoms and enhance mechanical stability. We also plan to improve the data rate of our GMOT LPAI by using a flat-top cooling beam with higher power to increase the initial atom number. Longer-term, the PIC components must see continued development, and hybrid integration strategies must be matured to realize a complete PIC-based laser system. We anticipate that a PIC laser system will significantly decrease cost and reduce manufacturing and testing time in realizing a complete matter-wave inertial measurement unit with three-axis atomic accelerometers and three-axis atomic gyroscopes, which will have tens of optical channels. The technologies developed in these ongoing efforts can also be applied beyond LPAI gravimeters, gravity gradiometers, accelerometers, and gyroscopes to other real-world atomic sensor applications, such as clocks, magnetometers, and electric field sensors.

\section*{Methods}

\subsection*{Grating Chip Fabrication}

\begin{figure}[b!]
    \centering
	\includegraphics[width=1\columnwidth]{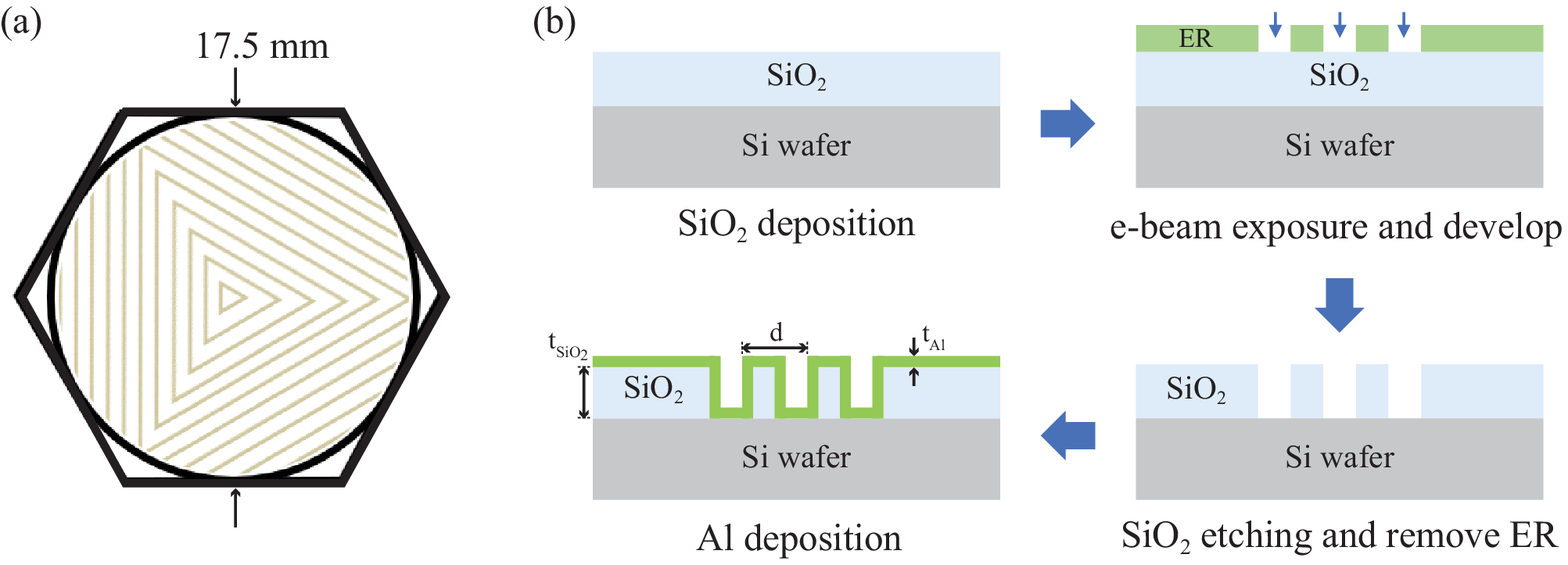}
	\caption{Grating design and fabrication process.
			(a) Schematic of the hexagonal grating chip showing a side-to-side length of \SI{17.5}{\mm} and three grating areas symmetrically arranged in the circular shape.
			The non-grating flat edges are used for the grating retainer clips. 
			The $1/e^2$ beam diameter of the cooling beam is \SI{28}{\mm} truncated to \SI{\approx 19}{\mm} by the fused silica viewport.	
			(b) Fabrication process of the GMOT grating chip.
			Parameters for the grating chip used in this paper are $\rm{t_\text{\ce{Al}}} = \SI{50}{\nm}$, $\rm{t_\text{\ce{SiO2}}} = \SI{195}{\nm}$, $\rm{d} = \SI{1.2}{\um}$, and \SI{\sim 50}{\percent} duty cycle (details in main text).
		}
	\label{fig:grating_chip_fabrication}
\end{figure}

The fabrication process is shown in \cref{fig:grating_chip_fabrication}(b).
A \SI{195}{\nm} layer of \ce{SiO2} was grown on a silicon wafer followed by spin-coating and soft-baking the e-beam resist (ER).
The 1D binary grating was then patterned in the ER via e-beam lithography.
The pattern was etched into the \ce{SiO2} using a selective reactive ion etching (RIE) process, stopping on the silicon. 
The ER was then removed with solvents and all remaining organic residue was removed in an \ce{O2} plasma. 
A \SI{50}{\nm} reflective coating of aluminum (\ce{Al}) was deposited conformally over the etched \ce{SiO2} grating, using an angled and rotating platen in an e-beam evaporator.
We chose Al over Au because the grating is necessarily inside the vacuum, where Rb can interact with the Au and substantially reduce its reflectivity \cite{Schwindt03}, and \ce{Al} gratings have been tested with \ce{Rb} atoms inside the vacuum \cite{McGilliganJOSAB2016, CotterAPB2016}.
The e-beam bias and the added sidewall thickness were carefully controlled to minimize errors in the target pitch, duty cycle, and depth of the grating.

\subsection*{Raman Beam Setup}

\begin{figure}[b!]
	\centering
	\includegraphics[width=0.7\columnwidth]{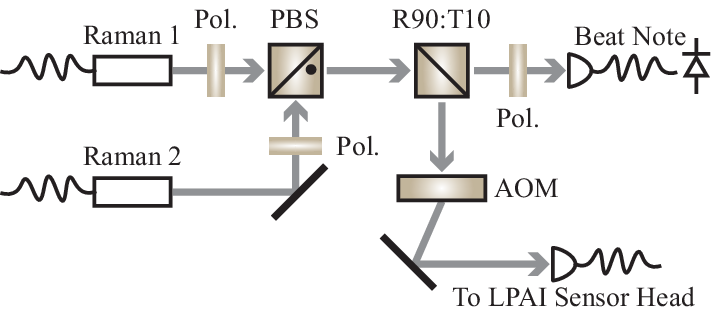}
	\caption{
			Optical setup for combining the two \SI{780}{\nm} Raman channels with crossed-linear-polarization and phase locking.
			Approximately \SI{10}{\percent} of the combined power is diverted for phase locking while the remaining power is directed towards an acousto-optic modulator for pulsing. 
			The combined Raman beams are coupled into the same polarization maintaining fiber with the polarizations aligned to the slow and fast axes of the fiber. PBS, polarizing beam splitter; R90:T10, non-polarizing cube beam splitter; Pol., polarizer; AOM, acousto-optic modulator.}
	\label{fig:raman_combiner}
\end{figure}

For the Raman channels, time-multiplexed frequency shifting can also be implemented, but was not pursued here in order to simplify phase stabilization. Both Raman beams were combined with orthogonal polarization using free-space optics, as shown in \cref{fig:raman_combiner}. A sample of the combined beams (\SI{\sim 10}{\percent}) was used for continuous Raman phase stabilization through the fiber-coupled \SI{1560}{\nm} AOM (or SSBM) and a voltage-controlled oscillator, while the majority of the power was sent to a free-space AOM for optical switching. The combined Raman light was then coupled to a single PM fiber along its two birefringent axes for delivery to the LPAI sensor head, reducing noise from varying paths.

%\section*{Data availability}
%All relevant data are available from the authors upon reasonable request.

%\section*{Code availability}
%Simulation code is available from the authors upon reasonable request.

% Bibliography

\section*{Acknowledgments}
This work was supported by the Laboratory Directed Research and Development program at Sandia National Laboratories. We thank Daniel Soh, Connor Brasher, Joseph Berg, Tony G. Smith, Greg Hoth, Bethany Little, Dennis J. De Smet, Melissa Revelle, Andrew L. Starbuck, Christina Dallo, Andrew T. Pomerene, Douglas Trotter, Christopher T. DeRose, Erik J. Skogen, Allen Vawter, Matt Eichenfield, Aleem Siddiqui, Peter Rakich, and Shayan Mookherjea  for their support and helpful discussion. Sandia National Laboratories is a multimission laboratory managed and operated by National Technology \& Engineering Solutions of Sandia, LLC, a wholly owned subsidiary of Honeywell International Inc., for the U.S. Department of Energy's National Nuclear Security Administration under contract DE-NA0003525.
This paper describes objective technical results and analysis. 
Any subjective views or opinions that might be expressed in the paper do not necessarily represent the views of the U.S. Department of Energy or the United States Government.

\section*{Author contributions}
J.L. and R.D. designed the LPAI experiments. J.L., R.D., and J.C. carried out the LPAI experiments. J. L., R.D., and P.D.D.S. analyzed the results. J.C., R.R.R., H.M., and G.B. supported the experiment. A.I., D.P.G., D.B., K.H.F., C.A.W., G.B., and J.L. developed the compact sensor head with the GMOT. J.L., P.S.F., J.R.W., W.K., and S.A.K. designed, fabricated, and tested the microfabricated grating-chip. A.L., M.G., A.K., G.B., and J.L. investigated the PIC laser architecture. J.L., R.D, and P.D.D.S prepared the manuscript for publication. P.D.D.S and J.L. coordinated the project.

%\section*{Competing interests} 
%The authors declare no competing interest.

\end{document}